\documentclass[conference]{IEEEtran}

\usepackage{cite}
\usepackage{amsmath,amssymb,amsfonts}
\usepackage{graphicx}
\usepackage{multirow}
\usepackage{textcomp}
\usepackage{xcolor}
\usepackage{balance}
\usepackage{stfloats}
\def\BibTeX{{\rm B\kern-.05em{\sc i\kern-.025em b}\kern-.08em T\kern-.1667em\lower.7ex\hbox{E}\kern-.125emX}}
\begin{document}

\title{Agent-Warden: eBPF-Based Kernel-Native Process-File Provenance Tracking for LLM Agents}

\author{
    \IEEEauthorblockN{
        Dongxu Cui\textsuperscript{1,2},
        Zhichao Gu\textsuperscript{2},
        Ping Zheng\textsuperscript{2},
        Simeng Han\textsuperscript{2}, and
        Yong Liao\textsuperscript{1,*}}
    \IEEEauthorblockA{
        \textsuperscript{1}\textit{School of Cyber Science and Technology, University of Science and Technology of China}, Hefei, China\\
        cdx@mail.ustc.edu.cn, yliao@ustc.edu.cn\\
        ORCID: 0009-0007-1008-5465 (D. Cui), 0000-0001-6403-0557 (Y. Liao)}
    \IEEEauthorblockA{
        \textsuperscript{2}\textit{China Greatwall Technology Group Co., Ltd.}, Shenzhen, China\\
        \{guzhichao, zhengping, hansimeng\}@greatwall.com.cn\\
        \textsuperscript{*}Corresponding author: Yong Liao (yliao@ustc.edu.cn)}
}

\maketitle

\begin{figure*}[!b]
\footnotesize
This is the author's version of a paper accepted for publication in the
Proceedings of the 2026 IEEE International Conference on Trust, Privacy
and Security in Intelligent Systems and Applications (IEEE TPS 2026).\\[3pt]
\copyright~2026 IEEE. Personal use of this material is permitted.
Permission from IEEE must be obtained for all other uses, in any current
or future media, including reprinting/republishing this material for
advertising or promotional purposes, creating new collective works, for
resale or redistribution to servers or lists, or reuse of any copyrighted
component of this work in other works.
\end{figure*}

\begin{abstract}
LLM agents execute dynamically generated process and file operations that are often invisible to application-layer tracing. We present Agent-Warden, 
an extended Berkeley Packet Filter (eBPF)-based provenance monitor for tracking task and regular-file states across process creation, file access, and process termination. 
Agent-Warden provides two interchangeable state backends: a PID-keyed hash-map backend for compatible kernels lacking BPF local-storage support and a task/inode-local-storage backend that couples 
state reclamation to kernel-object lifetimes. The system emits incremental causal edges to user space for asynchronous graph 
reconstruction and applies conservative exit-triggered causal aggregation to preserve causal context for short-lived proxy tasks. In a controlled 
file-mediated propagation scenario, Agent-Warden reconstructed a cross-process causal chain that was absent from the application-layer trace. 
On x86-64 and ARM64 bare-metal hosts, the evaluated workloads showed 0.2--3.5\% end-to-end overhead and 0.6--3.7\% additional system CPU time. 
These results indicate that the prototype provides kernel-level visibility with the measured overheads in the evaluated settings.
\end{abstract}


\section{Introduction}

Large Language Models (LLMs) with tool-calling capabilities enable agents to plan tasks, delegate subtasks, and invoke external tools \cite{surveyLLM2024}. 
This capability also expands the security impact of model-generated decisions. Indirect prompt injection can redirect tool-using applications \cite{greshake2023}, 
while recent studies have examined autonomous exploitation and agent-mediated attacks involving security-sensitive tool use \cite{fang2024,darkllm2025,ASB}. 
When an agent is permitted to create processes or access files, a high-level instruction can induce a sequence of low-level operating-system (OS) operations that may not be explicitly represented in the agent's application-level execution record.

Application-layer observability platforms such as Langfuse collect application-defined traces of model calls, tool invocations, and related execution spans \cite{langfuse}. 
Such telemetry is useful for analyzing application-level behavior, but its visibility is bounded by the events exposed through application instrumentation. 
Prior work on system-level agent observability similarly identifies cross-process execution as a reason to correlate application context with kernel events \cite{AgentSight2025}. 
Dynamically created scripts, native system calls, and asynchronous child processes may therefore produce OS activity that is absent from an application-layer trace.

At the kernel layer, process-associated state can be maintained in PID-keyed eBPF maps \cite{linuxBPFDocs,gregg2019bpf}. 
This design requires explicit state reclamation and careful handling of PID reuse; shared-map access may also introduce synchronization overhead under concurrent workloads. 
These considerations motivate examining an object-lifetime-coupled alternative. More broadly, the visibility boundary of application instrumentation motivates an independent kernel-level observability substrate that associates process and inode state and reconstructs process--file causality outside application-controlled telemetry.

To address this observability gap, we present Agent-Warden, an eBPF-based, kernel-native provenance tracing framework. 
To accommodate the heterogeneity of hardware compute capabilities and kernel versions in industrial deployments, Agent-Warden
introduces a dual-backend adaptive tracing architecture. 
The Warden-Hash engine uses a PID-keyed BPF hash map and provides expected constant-time lookup under typical workloads, supporting compatible kernels that lack BPF local-storage features.
The Warden-Local engine uses BPF task-local storage to associate provenance state with the kernel's task structure. 
This avoids repeated PID-keyed global map lookups on the common path and delegates state cleanup to the lifecycle management of the corresponding kernel object.

We formalize the tracked provenance transitions using a four-rule state-transition model over the provenance graph 
$G=(V,E)$. The model defines deterministic propagation for the process-derivation and file-interaction events considered in this work. 
Building upon this model, we introduce exit-triggered causal aggregation, which conservatively associates a parent process with the final provenance-influence state of a terminating child process. 
This mechanism preserves causal context for short-lived proxy tasks without claiming strict byte-level data flow.

The main contributions of this paper are summarized as follows:

\begin{itemize}
    \item \textbf{Dual-Backend State Management:} We design an eBPF-based provenance architecture featuring a PID-keyed hash backend for compatibility with kernels lacking BPF local-storage support 
    and a task/inode-local-storage backend for kernel-object-coupled state management.
    \item \textbf{Process--File Propagation Semantics:} We define and implement event-level causal propagation rules encompassing process derivation, regular-file operations, 
    namespace alterations (rename), and conservative exit-triggered causal aggregation, effectively addressing the asynchronous execution patterns of agents.
    \item \textbf{Cross-Architecture Evaluation:} We deploy and evaluate the prototype on x86-64 and ARM64 bare-metal systems, examining cross-temporal file-mediated propagation and the runtime overheads 
    of both state backends under bursty workloads.
\end{itemize}

\section{Background and Related Work}
\subsection{eBPF Execution and Security Model}\label{sec:ebpf-background}
The eBPF architecture defines a virtual-machine instruction set and an event-driven execution environment within the Linux kernel \cite{linuxBPFDocs,gregg2019bpf,sun2024verifier}. 
An eBPF program is attached to a supported kernel hook, such as a tracepoint, an fentry/fexit site, or a Linux Security Module (LSM) hook, and is invoked when the corresponding event occurs. 
Before a program is loaded, the kernel verifier analyzes it and rejects programs for which it cannot establish properties such as bounded control flow, valid memory accesses, and type-safe use of kernel helpers. 
Programs accepted by the verifier can be interpreted or compiled by a just-in-time compiler into native instructions. Program loading and attachment are additionally subject to kernel privilege and authorization controls.

eBPF maps provide persistent state across event invocations, while task and inode local-storage maps associate state with the lifetime of the corresponding kernel object. 
Ring buffers support asynchronous delivery of event records from kernel space to user space. Agent-Warden uses these mechanisms to update provenance state at kernel hooks and to deliver causal edges for user-space graph reconstruction. 
These mechanisms constrain the behavior of loaded monitoring programs but do not establish an absolute security boundary. Consistent with our threat model, we assume that the Linux kernel, the eBPF verifier and runtime, and the privileged component that loads Agent-Warden remain trusted.

\subsection{LLM Agent Security and Application-Layer Observability}\label{sec:related-llm}
Tool-using LLM agents can initiate security-sensitive system actions \cite{surveyLLM2024}. Prior studies have demonstrated indirect prompt-injection and agent-mediated attack scenarios in which unintended instructions lead to system effects \cite{darkllm2025,fang2024,greshake2023,ASB}. 
The OWASP Foundation also identifies excessive agency as a relevant risk for agentic applications \cite{owasp2025}. Existing mechanisms include declarative sandbox policies in AgentBound \cite{Buhler2026AgentBound}, application-layer tool-invocation rules in AgentSpec \cite{wang2026agentspec}, and application-level trace observability in Langfuse \cite{langfuse}.
These mechanisms have different enforcement and observation boundaries. Application-layer trace platforms observe events exposed by participating components, whereas permission and sandbox mechanisms enforce policies within their configured scope. 
Operations performed by dynamically generated scripts, native binaries, or uninstrumented child processes may therefore be absent from the application trace. Kernel-level observation provides a complementary source of process and file events outside this instrumentation boundary.

\subsection{System-Level Provenance Tracking and Bottlenecks of Traditional Auditing}\label{sec:related-provenance}
System-level provenance systems construct event-causality graphs from operating-system activity. Early approaches relied primarily on audit subsystems such as Linux Auditd. 
Although these methods can capture process-level information flows, systems such as ProTracer identify context switching and kernel-to-user-space data transfer as relevant sources of overhead \cite{king2003backtracking,Ma2016ProTracerTP}. 
At high event rates, these costs can affect collection throughput. EAudit and TAPAS address event loss and online-analysis overhead through buffering and graph partitioning, respectively \cite{10646884,309640,ma2017mpi}; however, event-capture and analysis costs remain design considerations for low-latency provenance monitoring \cite{li2021survey}.

\subsection{eBPF-Based Agent Observability and Comparison with AgentSight}\label{sec:related-ebpf}

AgentSight \cite{AgentSight2025} introduces boundary tracing to correlate an agent's high-level intent with its system-level effects. 
It uses eBPF probes to observe decrypted LLM communications and kernel events, and its user-space engine combines process lineage, temporal proximity, and argument matching before applying secondary LLM-based semantic analysis. 
Its primary objective is therefore to bridge the semantic gap between LLM interactions and system actions.

Agent-Warden addresses a narrower and complementary question: how process--file provenance state can be propagated and maintained when a causal path passes through a persistent file and may later continue in a process outside the originating process tree. 
It does not recover LLM intent or perform semantic threat interpretation. Instead, it defines explicit transitions for successful process and regular-file events, including process derivation, process-to-file propagation, file-to-process propagation, rename-associated namespace changes, and conservative exit-triggered aggregation. 
It also examines two kernel-state backends, including task/inode-local storage that couples provenance-state reclamation to kernel-object lifetimes.

Table~\ref{tab:agentsight-comparison} formalizes these distinctions. The comparison indicates that the contribution of Agent-Warden is not the use of eBPF itself, but the explicit process--file propagation semantics and the associated kernel-state-lifetime design.

\begin{table*}[t]
\centering
\caption{Mechanism-level comparison between AgentSight and Agent-Warden. ``Not specified'' indicates that the cited AgentSight paper does not describe the corresponding mechanism as part of its design; it does not imply that such a mechanism cannot be added.}
\label{tab:agentsight-comparison}
\footnotesize
\renewcommand{\arraystretch}{1.1}
\begin{tabular}{@{}p{0.17\textwidth}p{0.37\textwidth}p{0.39\textwidth}@{}}
\hline
\textbf{Dimension} & \textbf{AgentSight \cite{AgentSight2025}} & \textbf{Agent-Warden} \\
\hline
Primary objective &
Correlate LLM-level intent with system-level effects. &
Maintain kernel-level process--file provenance across synchronous and asynchronous execution. \\

Observed evidence &
Decrypted LLM traffic, process and system-call events, file activity, and network activity. &
Successful process-lifecycle events, regular-file reads and writes, and rename-associated namespace changes. \\

Causal mechanism &
Multi-signal correlation using process lineage, temporal proximity, argument matching, and secondary LLM analysis. &
Four explicit event-level propagation rules without LLM-based inference. \\

File-mediated propagation &
File events are included in correlated agent traces; explicit propagation from a written file to a later independent reader is not specified. &
Successful writes propagate process state to an inode, and successful reads propagate inode state to the reading process. \\

Kernel-state lifecycle &
Uses in-kernel filtering and a user-space stateful process tree; object-lifetime-coupled provenance storage is not a stated design objective. &
Provides a PID-keyed hash backend and a task/inode-local-storage backend coupled to kernel-object lifetimes. \\

Semantic scope &
Includes LLM intent and semantic interpretation of correlated traces. &
Intentionally limited to conservative OS-level causal association without inferring prompt intent. \\
\hline
\end{tabular}
\end{table*}

\section{System Model}
To delineate the scope of the agent observability framework, this section defines the system assumptions and analyzes the evasion threats faced by current monitoring mechanisms.

\subsection{System Assumptions and Scenarios}\label{SAAS}
This work focuses on the autonomous execution of LLM agents on host machines, including both x86-64 and ARM64 physical nodes. From a security standpoint, 
we assume that the underlying Linux kernel (including the eBPF subsystem) and hardware are initially trusted \cite{sun2024verifier,he2023cross}, and that the adversary or a compromised agent 
has not yet gained kernel-level privileges. The application-layer agent framework is considered an untrusted domain; it may issue malicious low-level system 
calls due to prompt injection, hallucinations, or unintended behaviors of the framework itself (e.g., unauthorized side-channel probing, covert data 
exfiltration). Under these assumptions, the goal of the framework is to observe agent-associated process-lifecycle and regular-file interactions within the explicitly defined monitoring scope.

\subsection{Threat Model and Observability Blind Spots}\label{TMAOBS}
Unlike traditional binary exploitation, security threats in agent environments primarily manifest as attempts to evade the observability system itself. Current 
observability approaches, which heavily rely on application-layer logs (e.g., planning outputs, API call monitoring), face two critical blind spots when confronted with 
black-box low-level execution:
\begin{itemize}
    \item A compromised agent can spawn and execute temporary scripts (e.g., stealthy data theft or state probing payloads) on a millisecond timescale, attempting 
    to complete operations and destroy the process before application-layer monitoring components can poll or read the state, thereby evading detection in macroscopic 
    system logs.
    \item Malicious payloads may cause the agent to create a child process group detached from the current terminal session to perform covert tasks. Once the critical 
    operation is dispatched, the intermediate proxy process exits immediately, thus severing the application-layer context between parent and child processes and 
    preventing the observability system from tracing backwards.
\end{itemize}

To address the above blind spots, the proposed dual-backend architecture aims to achieve three core observability goals: fine-grained lineage tracing, causal alignment 
across asynchronous execution cycles, and scope-bounded capture of process-lifecycle and regular-file events.

\section{Methodology}

\subsection{Dual-Backend Adaptive Observability Architecture}\label{M1}
This section describes the design and implementation of the Agent-Warden provenance tracing framework. Considering the 
heterogeneity of hardware capabilities and kernel versions in real-world deployments, Agent-Warden adopts a dual-backend 
adaptive architecture for underlying state maintenance, namely the Warden-Hash engine and the Warden-Local engine. The 
design aims to offer a pragmatic engineering trade-off between backward compatibility and kernel-object-coupled state management.

The Warden-Hash engine uses eBPF hash maps to maintain provenance state for system entities. 
These maps use mature kernel mechanisms, including Read-Copy-Update (RCU) \cite{mckenney2002rcu}, and provide average-case hash-table lookup. 
This backend is available on kernels that do not support BPF local storage. Under high task-creation rates or growth in retained state, the shared global map may still incur hash collisions and cross-core synchronization costs.

To provide kernel-object-coupled state management and per-object state separation, this work further introduces the Warden-Local engine. 
This engine uses BPF task-local storage to associate the required provenance marker with the kernel's \texttt{task\_struct} object. 
This binding isolates state across thread groups while allowing all threads within the same TGID to access a shared canonical provenance marker, avoiding repeated PID-keyed 
global map lookups on the common access path. More importantly, Warden-Local couples the lifecycle of the provenance state to that of the corresponding kernel object. 
When a transient task terminates, the kernel reclaims the object-associated state, so the probes do not require a separate deletion path for the corresponding task-local entry. Furthermore, 
to maintain unified kernel-object-coupled state management across all system entities, this local-storage paradigm is symmetrically applied to file operations 
via \texttt{BPF\_MAP\_TYPE\_INODE\_STORAGE}, ensuring that file provenance markers remain associated with the corresponding inode for the lifetime of that inode object.

\subsection{Causal Flow Transition Model Based on Boolean Matrix Algebra}\label{M2}
\begin{figure}[htbp]
\centering
\includegraphics[width=\linewidth]{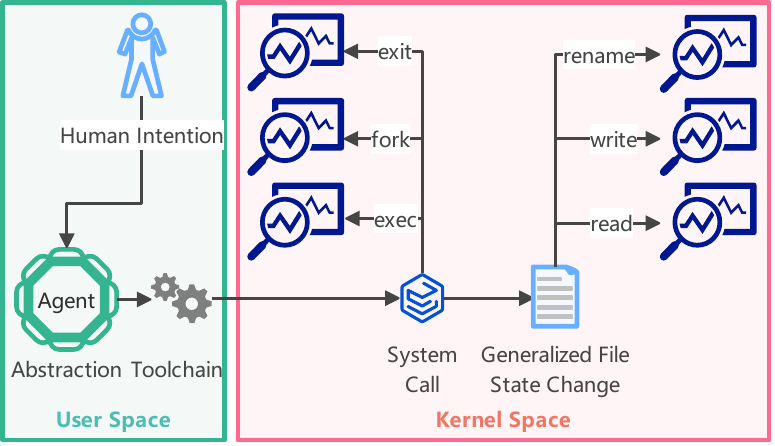}
\caption{The six atomic operations and the system call layer.}
\label{fig:agent-call}
\end{figure}

From the perspective of kernel interaction, any program's modification of system state can ultimately be reduced to ordered read 
and write operations on generalized files (i.e., state carriers). As illustrated in Figure~\ref{fig:agent-call}, regardless of whether a program 
manifests as 3D graphics rendering, neural network training, financial transaction processing, or autonomous agent execution, 
its interaction with the kernel can be decomposed into a finite combination of the following six atomic operations:
\begin{itemize}
    \item exec: replaces the current process image and creates a new process-version node in the provenance graph.
    \item fork: duplicates the computational context to extend parallelism.
    \item exit: terminates computation and returns results.
    \item read: retrieves data from persistent state.
    \item write: modifies persistent state.
    \item rename: updates the namespace mapping while preserving any provenance marker associated with the underlying inode, rather than introducing a new content-flow transition.
\end{itemize}

To eliminate false causal edges generated by denied or failed operations, Agent-Warden uses return-side hooks, such as \textit{kretprobes} or \textit{fexit} programs, 
to validate the success and transferred byte count of file operations. Process-lifecycle transitions are captured through dedicated kernel events, such 
as \texttt{sched\_process\_fork}, \texttt{sched\_process\_exec}, and \texttt{sched\_process\_exit}. State updates are performed only after the corresponding operation 
is known to have succeeded (e.g., $ret > 0$ for byte transfers in \textit{read}/\textit{write}, $ret == 0$ for \textit{rename}).

Based on this, Agent-Warden intercepts the aforementioned kernel atomic operations via eBPF hooks to achieve low-level provenance 
tracking. To formally describe the multi-dimensional provenance propagation mechanism within the framework, we define a four-rule state-transition model. 
Processes and files in the system are abstracted as graph nodes, and system calls are abstracted as
directed edges. Inspired by established information-flow tracking principles \cite{taintdroid2010}, Agent-Warden defines the following four provenance propagation rules:

\textbf{Process Derivation} $(P\rightarrow P)$: On fork or clone, the child inherits the provenance state of its parent. On exec, the current task retains its provenance 
state while the system records a version edge from the pre-exec process image to the newly loaded image. These rules keep agent-created tasks and program-image transitions 
within the same provenance lineage. This ensures that all subtasks spawned by an agent are included in
the provenance graph. To maintain process-wide provenance consistency across threads, our model enforces state consistency at the Thread Group ID (TGID) level. Architecturally, 
rather than maintaining isolated states per thread, Agent-Warden resolves the thread-group leader for each event and uses the leader's task-local storage as the canonical process-level state. 
Consequently, all threads in the same TGID access the same monotonic provenance marker. This avoids explicit replication of state across individual thread-local entries.

\textbf{Entity Propagation $(P \rightarrow F)$}: When a tracked process successfully writes data to a regular file, the process's provenance marker is propagated to the corresponding inode. 
A successful rename operation does not introduce a new content-flow transition; it is recorded as a namespace edge while preserving any marker already associated with the underlying inode.

\textbf{Read-Based Propagation ($F \rightarrow P$)}: When an unmarked process successfully reads data from a provenance-marked file, the file's provenance marker is propagated along the read 
edge to the process. This preserves the causal association when a previously unrelated process consumes data produced within the monitored agent lineage.

\textbf{Exit-Triggered Causal Aggregation} $(P\rightarrow P)$: Upon termination of a controlled child process, Agent-Warden records a conservative child-to-parent causal-association edge and updates 
the parent's provenance-influence state. This transition does not imply strict byte-level data flow. Instead, it intentionally over-approximates causal linkage so that short-lived proxy tasks remain 
associated with the agent's broader execution lineage.

To express the four-rule state-transition model in Boolean algebra, we define a causal flow transition model based on 
Boolean matrix algebra.

Let the global set of system entities be $V = P \cup F$, where $P$ is the set of processes and $F$ the set of files. At logical time $t$, the row vector 
$\mathbf{\tau}^{(t)} \in \mathbb{B}^{|V|}$ (where $\mathbb{B} = \{0, 1\}$) represents the conservative provenance-influence state of system entities. A marked 
entity (i.e., $\mathbf{\tau}_i^{(t)} = 1$) is causally associated with the monitored agent lineage; this association may arise from observed data transfer or 
conservative exit-time aggregation and therefore does not necessarily imply byte-level data dependence.

For any discrete, atomic low-level system event $e_t = \langle op, u, v \rangle$ occurring at logical time $t$ (where $u$ represents the active subject/initiator and $v$ represents 
the passive target), a causal flow projection matrix $C^{(t)} \in \mathbb{B}^{|V| \times |V|}$ is defined to map the operational semantics to a directed edge in the macro-level graph structure. 
Its formal definition is as follows:
\begin{align}
C^{(t)} &= (\mathbb{I}_{fork}(e_t) + \mathbb{I}_{exec}(e_t) + \mathbb{I}_{write}(e_t) + \mathbb{I}_{exit}(e_t)) M_{u,v} \nonumber \\
&\quad + \mathbb{I}_{read}(e_t) M_{v,u}
\end{align}

where $\mathbb{I}_{op}(e_t)$ is the event type indicator function, which takes the value $1$ if and only if the operation type of $e_t$ is $op$, and $\mathbf{M}_{i, j}$ is a basis matrix with 
a $1$ at position $(i,j)$ and $0$ elsewhere. For a \textit{read} event, the causal edge is oriented from the file object $v$ to the process $u$ because data flows from the passive object $v$ to 
the active subject $u$; therefore, the read term uses $\mathbf{M}_{v,u}$. Conversely, a \textit{write} event is oriented from $u$ to $v$. Note that the \textit{rename} operation is omitted 
from $C^{(t)}$ because it emits a namespace-change edge without modifying the provenance-influence state $\mathbf{\tau}^{(t)}$.

To initialize the system, all designated agent roots are explicitly tracked, i.e., $\mathbf{\tau}_{r}^{(0)}=1$ for $r\in A$, where $A$ is the set of monitored agent roots. Based on the projection matrix, 
the state transition of the global provenance graph is accomplished through deterministic Boolean vector algebra operations without the need for traversal or complex logical judgment. 
The unified state transition evolution equation takes the following simplified form:
\begin{equation}
\mathbf{\tau}^{(t+1)} = \mathbf{\tau}^{(t)} \lor \left(\mathbf{\tau}^{(t)} \otimes C^{(t)}\right)
\end{equation}

where $\otimes$ denotes Boolean matrix multiplication, and $\lor$ denotes the element-wise OR operation on vectors.

\subsection{Architectural Implementation and Asynchronous Reconstruction}
Equations (1) and (2) formally specify the causal semantics of low-level events, rather than mandating an implementation based 
on dense matrix operations. In our system architecture, constructing the global provenance graph does not rely on computing high-dimensional matrices. 
Instead, each observed kernel event is directly reduced to an expected constant-time state update on its specific source and destination entities via the eBPF mechanism.

To reconstruct the global provenance graph without stalling kernel execution, Agent-Warden adopts an ``in-kernel local scalar update, user-space asynchronous reconstruction'' paradigm. 
For each relevant successful event, including provenance-propagation and namespace-change events, the framework asynchronously emits an incremental causal edge to user space via an eBPF ring buffer. 
This design decouples graph reconstruction from the system-call path and avoids synchronous user-space I/O during event processing.

\subsection{Architectural Sources of Runtime Overhead}

Table~\ref{tab:overhead-sources} decomposes the runtime work introduced by Agent-Warden. 
The synchronous kernel path includes eBPF hook dispatch and program execution, event qualification, provenance-state access and update, and ring-buffer submission for emitted edges. 
User-space graph reconstruction is asynchronous with respect to the triggering system call but still contributes to aggregate system CPU consumption. 
State reclamation differs between the two backends: Warden-Hash performs explicit map-state maintenance, whereas Warden-Local delegates reclamation of local-storage entries to the lifetime of the associated kernel objects.

A static eBPF instruction count alone would not account for the runtime costs of map and local-storage helper calls, object allocation, synchronization, or ring-buffer delivery. 
We therefore use the architectural breakdown below to identify the relevant cost centers, while reporting only aggregate end-to-end and system CPU measurements in the evaluation.

\begin{table}[htbp]
\caption{Architectural Sources of Agent-Warden Runtime Overhead}
\label{tab:overhead-sources}
\centering
\scriptsize
\setlength{\tabcolsep}{2pt}
\begin{tabular}{@{}p{0.22\columnwidth}p{0.45\columnwidth}p{0.24\columnwidth}@{}}
\hline
Stage & Introduced work & Backend dependence \\
\hline
Hook execution &
Dispatch and execute the attached eBPF program for each candidate event. &
Shared by both backends. \\

Event qualification &
Identify the involved entities and validate operation success and transferred byte count. &
Shared by both backends. \\

State access and update &
Retrieve provenance markers and apply the corresponding scalar propagation rule. &
Hash-map helpers for Warden-Hash; task/inode-local-storage helpers for Warden-Local. \\

Edge emission &
Reserve, populate, and submit a ring-buffer record for each emitted causal or namespace edge. &
Shared by both backends. \\

State reclamation &
Remove or reclaim provenance state when the corresponding entity terminates. &
Explicit map maintenance for Warden-Hash; kernel-object-lifetime reclamation for Warden-Local. \\

Graph reconstruction &
Consume emitted records and update the provenance graph in user space. &
Shared and asynchronous. \\
\hline
\end{tabular}
\end{table}

\section{Experiment}

Our evaluation addresses two research questions. \textbf{RQ1} asks whether Agent-Warden can reconstruct a file-mediated causal chain that crosses the boundary of application instrumentation. 
\textbf{RQ2} examines the runtime overhead and cross-architecture behavior of the two kernel-state backends. 
RQ1 uses Langfuse as an application-layer functional visibility baseline under the same workload, whereas RQ2 compares each Agent-Warden backend with an uninstrumented execution.

\subsection{Experimental Setup and Workloads}\label{E1}

To evaluate the functionality and performance overhead of the Agent-Warden framework and its dual-backend provenance engines in 
heterogeneous hardware environments, we conduct experiments on two bare-metal deployment scenarios to exclude interference from 
virtualization:

\begin{itemize}
  \item \textbf{x86-64 General-Purpose Node}: Equipped with an AMD 9950X processor (16 cores, 32 threads).
  \item \textbf{ARM64 Edge Computing Node}: Equipped with a Phytium FT-D2000/4 processor (8 cores, 8 threads), used to evaluate 
  cross-ISA portability.
\end{itemize}

All test nodes run Ubuntu 24.04 LTS with Linux kernel 6.8. This kernel provides native support for features such as BPF Type Format (BTF) and BPF task-local storage. 
Monitoring probes are developed using Clang~18 and libbpf, and are implemented with BTF and CO-RE to facilitate portability across compatible kernel versions. 
Although the Warden-Hash engine is architecturally designed to provide backward compatibility for kernels lacking modern local storage features, our evaluation 
standardizes on Linux kernel 6.8 to provide a consistent basis for comparing the two state backends.

We use Langfuse as an application-layer functional visibility baseline rather than as a like-for-like performance baseline. 
Langfuse records traces emitted by instrumented application components and supports distributed tracing when participating components explicitly propagate a shared trace context \cite{langfuseObservabilityDocs,langfuseDistributedTracingDocs}. 
However, instrumenting the parent agent does not automatically instrument arbitrary child processes, dynamically generated scripts, or native binaries. 
Trace context is also not automatically propagated through a regular file to an independent process that later reads the file. 
Instrumenting every participating component would modify the evaluated workload and would not provide the same kernel-event stream of successful process, read, write, and rename operations. 
We therefore compare Langfuse and Agent-Warden in terms of the causal structures visible under the same unmodified workload, rather than comparing their runtime overheads as if they implemented an equivalent event scope.

At the kernel layer, generic eBPF security monitors such as Falco and Tetragon have different monitoring and enforcement objectives and do not implement the process--file propagation semantics evaluated here. 
A quantitative comparison would therefore combine different event scopes and security models. 
For RQ2, we measure Warden-Hash and Warden-Local against uninstrumented execution under identical workloads to characterize their respective runtime overheads.

\subsection{Causal Tracking Effectiveness and Cross-Boundary Provenance}
\label{sec:security_effectiveness}

\begin{figure}[htbp]
\centering
\includegraphics[width=\linewidth]{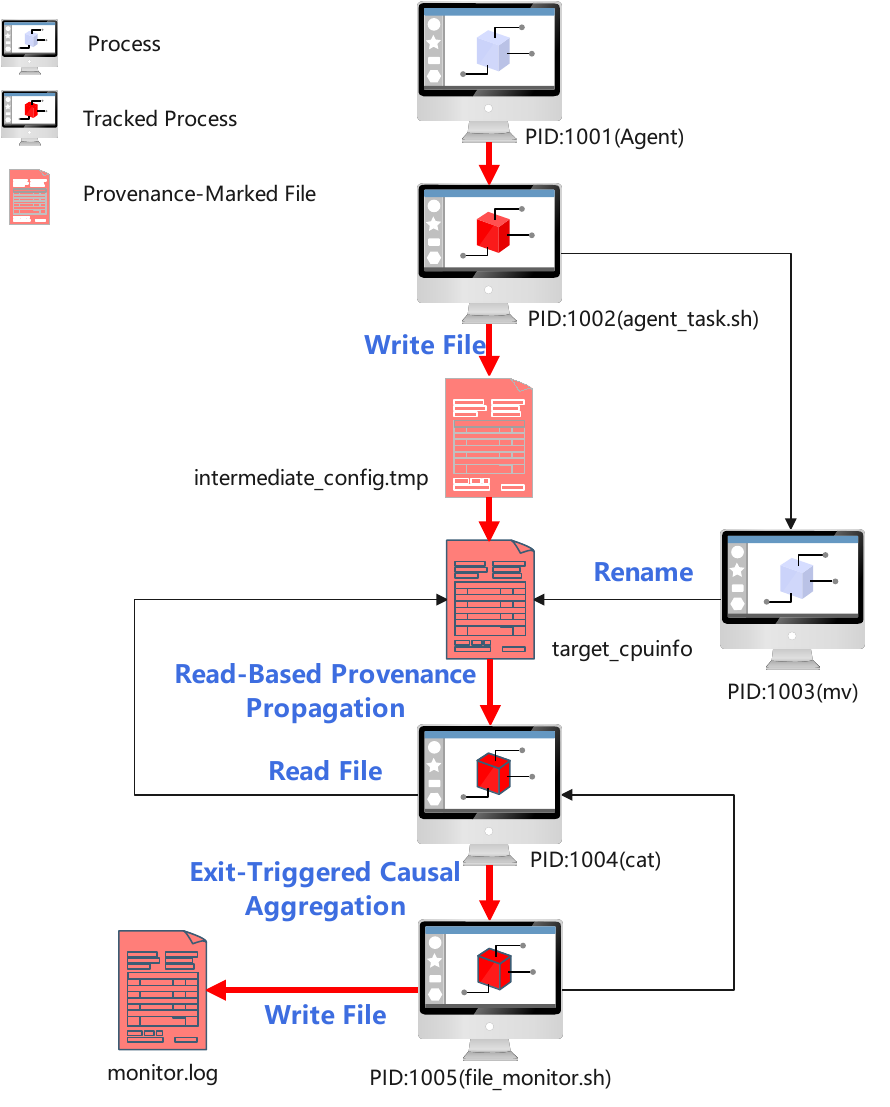}
\caption{Core causal subgraph of the file-mediated workload. PID 1003 records the rename from \texttt{intermediate\_config.tmp} to \texttt{target\_cpuinfo}; the operation preserves the inode-associated provenance marker.}
\label{fig:Infection-Graph}
\end{figure}

To examine the causal tracking capability of the framework in a controlled cross-boundary scenario, we use a predefined payload that generates the expected process--file propagation chain.
We provide the agent with a predefined adversarial test task involving cross-boundary interactions, which produces the expected core causal subgraph shown in Figure~\ref{fig:Infection-Graph} under Agent-Warden's conservative propagation semantics.

In this adversarial scenario, the agent process (PID 1001) spawns the task script \texttt{agent\_task.sh} (PID 1002), which writes the provenance-marked temporary file \texttt{intermediate\_config.tmp}. 
The task script then invokes \texttt{mv} (PID 1003), which successfully renames this file to \texttt{target\_cpuinfo}. 
Agent-Warden records this operation as a namespace-change edge while preserving the provenance marker associated with the underlying inode; the rename does not introduce an additional content-flow transition. 
Subsequently, the independently running and uninstrumented host script \texttt{file\_monitor.sh} (PID 1005) spawns \texttt{cat} (PID 1004), which reads the renamed file.

Under the same workload, Langfuse records the events emitted within the instrumented agent application but does not contain the subsequent operations of the uninstrumented host process. 
After the generated task script writes and renames the temporary file, no application-level trace context is carried through the renamed file to the independent monitor process. 
Consequently, the Langfuse trace does not reconstruct the file-to-process causal continuation or associate the later write to \texttt{monitor.log} with the originating agent trace. 
This result reflects the boundary of the selected application instrumentation rather than an inability of Langfuse to correlate components that explicitly propagate a shared trace context.

\begin{table}[htbp]
\caption{Functional Visibility under the File-Mediated Workload}
\label{tab:langfuse-comparison}
\centering
\scriptsize
\setlength{\tabcolsep}{3pt}
\begin{tabular}{@{}p{0.50\columnwidth}cc@{}}
\hline
Observed information & Langfuse & Agent-Warden \\
\hline
Instrumented application-level execution & Yes & Out of scope \\
OS process derivation events & Not recorded & Recorded \\
Successful regular-file read/write/rename events & Not recorded & Recorded \\
Causal continuation through a file to an independent reader & Not reconstructed & Reconstructed \\
Prompt and model-call semantics & Recorded & Out of scope \\
\hline
\end{tabular}
\end{table}

Compared with the application-layer trace, Agent-Warden records the causal structures defined by its four-rule state-transition model. As shown in the tracing topology in 
Figure~\ref{fig:Infection-Graph}, the system records process derivation ($P \rightarrow P$) and entity propagation ($P \rightarrow F$), and applies read-based propagation ($F \rightarrow P$) when 
an independent process reads the provenance-marked file. Moreover, upon the termination of this child process, the exit-triggered causal aggregation conservatively associates the parent process 
with the child's final provenance-influence state.

The reconstructed subgraph contains the expected cross-boundary causal chain defined by the propagation rules.

To assess repeatability beyond a single successful trace, we executed the end-to-end workload 30 times for each state backend on both the x86-64 and ARM64 platforms. 
After each run, we manually inspected the reconstructed provenance graph to verify the expected process-derivation behavior, process-to-file propagation, 
rename-edge recording with inode-marker preservation, file-to-process propagation, exit-triggered causal aggregation, and complete causal-chain reconstruction. 
As shown in Table~\ref{tab:functional-validation}, both backends reproduced all expected behaviors in all 30 runs on each platform.

\begin{table}[htbp]
\caption{Repeatability of Expected Propagation Behaviors on x86-64 and ARM64 Platforms (30 Runs per Backend per Platform)}
\label{tab:functional-validation}
\centering
\scriptsize
\setlength{\tabcolsep}{3pt}
\begin{tabular}{@{}lcc@{}}
\hline
Behavior & Warden-Hash & Warden-Local \\
\hline
Process derivation and exec continuity & 30/30 & 30/30 \\
Process-to-file propagation            & 30/30 & 30/30 \\
Rename edge and marker preservation    & 30/30 & 30/30 \\
File-to-process propagation            & 30/30 & 30/30 \\
Exit-triggered causal aggregation      & 30/30 & 30/30 \\
Complete causal-chain reconstruction   & 30/30 & 30/30 \\
\hline
\end{tabular}
\end{table}

\begin{figure}[htbp]
\centering
\includegraphics[width=\linewidth]{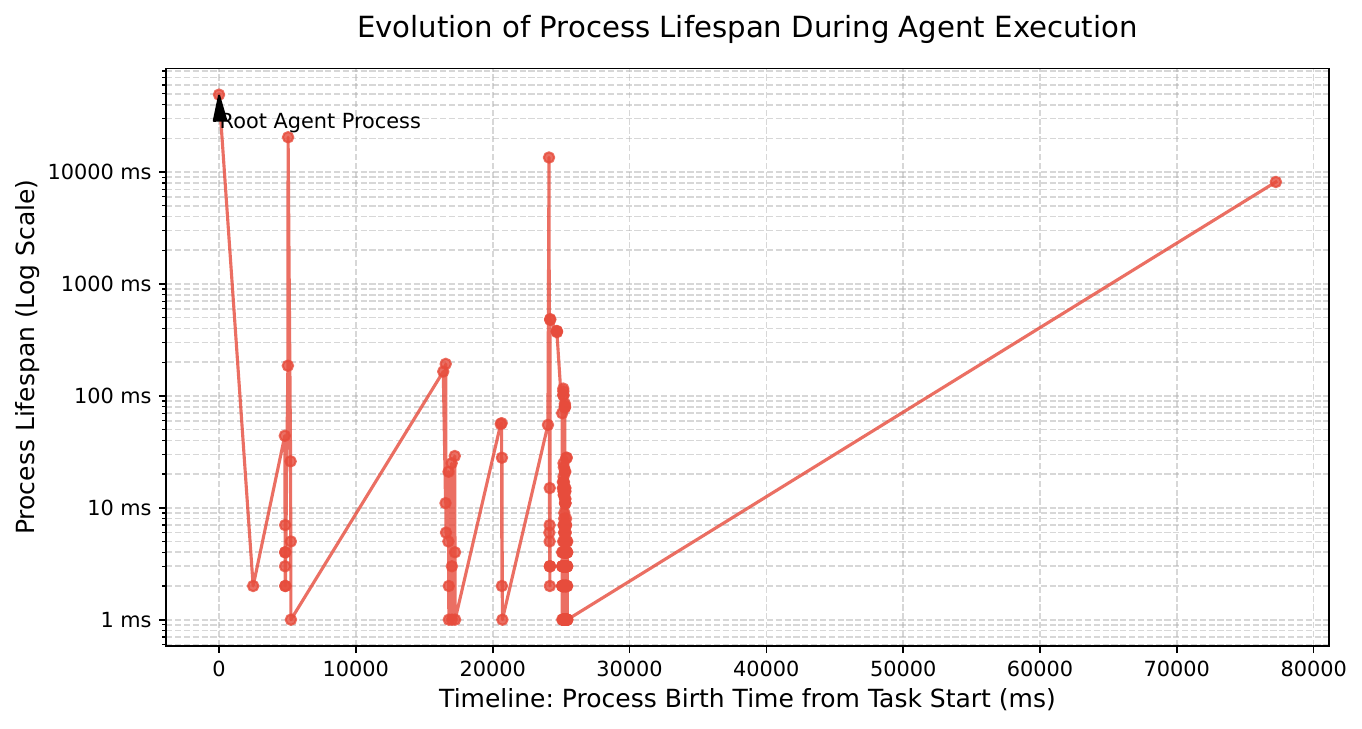}
\caption{Temporal distribution and diverse lifespans of proxy tasks spawned by the LLM agent workload. This asynchronous execution pattern illustrates the need for object-lifetime-coupled provenance 
tracking through file-mediated causality.}
\label{fig:lifespan-evolution}
\end{figure}

Figure~\ref{fig:lifespan-evolution} illustrates the cross-temporal behavior observed in the evaluated workload. 
The scatter plot shows the temporal distribution and diverse lifespans of proxy tasks, motivating cross-temporal, object-lifetime-coupled tracking.
Near the end of the time axis (around $t = 77,000\text{ ms}$), the system records a trajectory point that is temporally separated 
from the main task flow. This event corresponds to an independent asynchronous polling service on the host. Tens of thousands of milliseconds 
after the agent's core task flow had terminated, this polling process attempted to read an intermediate configuration file carrying the agent-associated provenance marker. 
Application-context-bound tracing may lose this association once the original agent execution context terminates. However, Agent-Warden 
successfully retained the causal markers on the corresponding inode object, capturing this asynchronous causal association.

Because provenance state is maintained at the kernel layer and coupled to the inode lifetime, 
Agent-Warden associated the delayed read with the earlier agent-generated file state even after the original application-layer context had terminated. 
This result demonstrates cross-temporal tracking within the lifetime of the corresponding in-memory inode object.

\subsection{Stress Testing under Bursty Workloads}
\label{sec:stress_test}

To quantify the runtime overhead of Agent-Warden under a bursty process-and-file workload, we construct a synthetic ``agent behavior primitive cycle.''
The workload repeats a sequence of process creation (fork/exec), command invocation, file read/write, and rename operations 1,000 times to generate a bursty kernel-event workload resembling an agent task.

Three control baselines are established: a native system without probes (Baseline Off), the Agent-Warden hash-map-based tracking 
engine (Warden-Hash), and the Agent-Warden local-storage-based tracking engine (Warden-Local). 
The experiments are conducted separately on physical x86-64 and ARM64 nodes.

Across the evaluated workloads, the measured end-to-end overhead was 0.2--3.5\%, and the additional system CPU time was 0.6--3.7\%. 
Table~\ref{tab:overhead} summarizes these measurements relative to the baseline system across both platforms 
(mean $\pm$ standard deviation over 90 measurement rounds).

\begin{table}[htbp]
\centering
\caption{Performance Overhead of Agent-Warden Relative to Baseline (mean $\pm$ std, $n=90$ rounds)}
\label{tab:overhead}
\begin{tabular}{|l|c|c|c|}
\hline
\textbf{Platform} & \textbf{Engine} & \textbf{End-to-End (\%)} & \textbf{System CPU $\Delta$ (\%)} \\
\hline
\multirow{2}{*}{x86-64} & Warden-Hash  & $0.2 \pm 1.2$ & $0.6 \pm 3.1$ \\
                         & Warden-Local & $1.1 \pm 1.4$ & $2.5 \pm 3.3$ \\
\hline
\multirow{2}{*}{ARM64} & Warden-Hash  & $3.5 \pm 0.4$ & $3.7 \pm 0.8$ \\
                        & Warden-Local & $3.5 \pm 0.4$ & $3.7 \pm 0.8$ \\
\hline
\end{tabular}
\end{table}

The larger relative variance observed on x86-64 may partly reflect its smaller absolute baseline latency (1.57\,s versus 8.30\,s on ARM64), because a similar absolute timing fluctuation produces a larger percentage change when the baseline is shorter.

The reported end-to-end and system CPU measurements aggregate the cost centers summarized in Table~\ref{tab:overhead-sources}. 
Both backends incur hook-dispatch, event-qualification, scalar state-update, ring-buffer, and asynchronous reconstruction costs. 
Their backend-specific difference lies primarily in state access and reclamation: Warden-Hash uses key construction and hash-map helper operations together with explicit state maintenance, whereas Warden-Local uses task/inode-local-storage helpers and kernel-object-coupled reclamation. 
Local storage therefore changes the state-management path but does not imply lower aggregate CPU consumption for every workload.

Under the evaluated x86-64 workload, Warden-Local has a higher measured system CPU delta than Warden-Hash, while the two backends have the same rounded overhead values on ARM64. 
Because the current measurements do not isolate individual stages, these differences cannot be attributed to a single helper, map operation, or event type. 
The results support only the aggregate conclusion that both backends have comparable end-to-end overhead within the evaluated workloads. 
Warden-Hash provides compatibility with kernels lacking local storage, whereas Warden-Local provides object-lifetime-coupled state management.

\section{Discussion}
\label{sec:discussion}

\paragraph{Interpretation of Experimental Results}

Within the evaluated workloads, both Agent-Warden backends reproduced the expected four-rule process and file propagation behavior. 
Notably, exit-triggered causal aggregation preserves a conservative causal association between short-lived child processes and their parent context beyond the application-layer trace boundary.
Warden-Local uses task/inode-local storage and reduces reliance on PID-keyed global maps, whereas Warden-Hash supports kernels that lack BPF local-storage features. 
Because the measurements aggregate multiple cost centers, the observed overhead differences cannot be attributed to a single state-access mechanism.

\paragraph{Comparative Analysis with Existing Work}

As summarized in Table~\ref{tab:agentsight-comparison}, AgentSight and Agent-Warden both use eBPF for system-level observation but address different causal questions. 
AgentSight correlates LLM communications with system effects and applies semantic analysis, whereas Agent-Warden does not infer high-level intent. 
Agent-Warden instead focuses on explicit process--file state propagation, including the continuation of provenance through a persistent file to a later reader, and on kernel-object-coupled provenance-state management. 
The two systems should therefore be viewed as complementary rather than as interchangeable implementations of the same observability model.

Langfuse is used only as an application-layer visibility reference rather than as a performance baseline. 
Runtime overhead is evaluated by comparing Warden-Hash and Warden-Local with the uninstrumented system under identical workloads. 
Functional behavior is examined against a predefined expected subgraph generated by the controlled payload.

\paragraph{Limitations}

Within the evaluated process--file workloads, the prototype reproduced the expected causal structures with the measured overheads; its current scope has the following limitations:

\begin{itemize}
    \item \textbf{Observability Scope and Granularity:} The current state-transition model primarily captures macro-level process lifecycles and standard file I/O operations. 
    Advanced kernel interactions (e.g., asynchronous \texttt{io\_uring} events, memory-mapped I/O) and covert side-channel communications (e.g., timing characteristics 
    via \texttt{nanosleep}) remain invisible. Extending observability to these granular events is targeted for future architectural iterations.
    \item \textbf{Causal Precision and Semantic Gap:} The framework constructs physical flow topology at the system call level, lacking the semantic context of application-layer prompts. 
    Furthermore, the conservative exit-triggered aggregation explicitly prioritizes causal linkage over strict data dependence, which may introduce causal over-approximation (false positives) 
    if a parent process does not explicitly consume the proxy child's output. Additionally, process reparenting may cause exit-time aggregation to occasionally refer to a subreaper (e.g., systemd) 
    rather than the original spawning process, a boundary case designated for future refinement.
    \item \textbf{State Persistence Volatility:} To maintain expected constant-time in-memory state access, inode-local labels are currently coupled to the lifetime of in-memory inode objects. 
    They are not persistent across severe memory pressure (inode cache reclamation) or system reboots. Non-volatile tracking could be implemented using filesystem extended attributes or an external persistent store, 
    at the cost of additional metadata-management and persistence overhead; this extension is left to future work.
\end{itemize}

\paragraph{Future Work}

To address the above limitations, future work will proceed along the following directions:

\begin{enumerate}
  \item \textbf{Cross-layer semantic alignment and end-to-end attribution:} Explore a joint observability architecture that 
  correlates the kernel-level provenance graph with application-layer processing logs along temporal and entity-level dimensions, in order 
  to bridge the semantic gap between low-level system actions and natural language instructions.
  \item \textbf{Fine-grained tracing and side-channel detection:} Extend the observation scope to memory-mapped I/O, \texttt{io\_uring}, pipes, 
  Unix-domain sockets, shared-memory IPC, and network endpoints, and introduce heuristic anomaly detection methods targeting timing-related system 
  calls to mitigate the risk of side-channel data exfiltration. Meanwhile, refine the provenance propagation logic for file I/O to improve the precision of inferred data-flow directions.
  \item \textbf{Active intervention and defense mechanisms:} Investigate the feasibility of extending the framework with active policy enforcement. Explore active enforcement 
  through BPF LSM hooks or integration with Linux security mechanisms such as seccomp, enabling policy-based denial or termination when a prohibited provenance transition is detected.
  \item \textbf{Component-Level Performance Profiling:} The current evaluation reports aggregate end-to-end latency and system CPU consumption. 
  Table~\ref{tab:overhead-sources} identifies the architectural cost centers but does not quantify their individual contributions. 
  Future work will separately measure per-hook execution, state-helper operations, ring-buffer delivery, and user-space reconstruction.
\end{enumerate}

\section{Conclusion}
\label{sec:conclusion}

This paper presents Agent-Warden, a kernel-level provenance tracking framework designed to address observability gaps in the process and file interactions of LLM agents.
The four-rule state-transition model enables Agent-Warden to capture event-level process creation, file-interaction behavior, and asynchronous execution paths. 
Exit-triggered causal aggregation preserves conservative causal context for short-lived child processes whose execution may extend beyond the application-layer trace boundary.

In terms of engineering implementation, Agent-Warden provides two interchangeable tracking engines. Warden-Hash supports compatible kernels lacking BPF local-storage features, 
while Warden-Local couples provenance-state management to task and inode lifetimes and reduces reliance on PID-keyed global maps. 
Experiments on x86-64 and ARM64 bare-metal platforms show that Agent-Warden reconstructs the expected causal chain in the evaluated scenario while incurring 0.2--3.5\% end-to-end overhead 
and 0.6--3.7\% additional system CPU time. Together, the functional and performance results support the feasibility of kernel-level process--file provenance tracking for LLM agents within the evaluated workloads.

\section*{AI Disclosure}

We used OpenAI Codex, DeepSeek, and Google Gemini to assist with English-language editing, manuscript organization, and consistency-oriented revision. The tools materially affected the Abstract and Sections I, II, IV, V, VI, and VII. The authors verified the correctness and originality of all content including references.

\bibliographystyle{IEEEtran}
\balance
\bibliography{agent-warden}
\end{document}